# COSIMA-Rosetta calibration for in-situ characterization of 67P/Churyumov-Gerasimenko cometary inorganic compounds.

Harald Krüger[1], Thomas Stephan[2], Cécile Engrand*[3], Christelle Briois[4], Sandra Siljeström[5], Sihane Merouane[1], Donia Baklouti[6], Henning Fischer[1], Nicolas Fray[7], Klaus Hornung[8], Harry Lehto[9], François-Régis Orthous-Daunay[10], Jouni Rynö[11], Rita Schulz[12], Johan Silen[11], Laurent Thirkell[4], Mario Trieloff[13], Martin Hilchenbach[1].

[1]Max-Planck-Institut für Sonnensystemforschung, Justus-von-Liebig-Weg 3, 37077 Göttingen, Germany
[2]Department of the Geophysical Sciences, The University of Chicago, 5734 S Ellis Ave, Chicago, IL 60637, USA
[3]Centre de Sciences Nucléaires et de Sciences de la Matière, CNRS/IN2P3-Univ. Paris Sud - UMR8609, Batiment 104, 91405 Orsay campus, France.
[4]Laboratoire de Physique et Chimie de l'Environnement et de l'Espace (LPC2E), CNRS/ Université d'Orléans, 45071 Orléans, France
[5]Department of Chemistry, Materials and Surfaces, SP Technical Research Institute of Sweden, Box 857, 50115 Borås, Sweden,
[6]Institut d'Astrophysique Spatiale, CNRS / Université Paris Sud, Orsay Cedex, France
[7]LISA, UMR CNRS 7583, Université Paris Est Créteil et Université Paris Diderot, Institut Pierre Simon Laplace, France
[8]Universität der Bundeswehr LRT-7, Werner Heisenberg Weg 39, 85577 Neubiberg, Germany
[9]University of Turku, Department of Physics and Astronomy, Tuorla Observatory Väisäläntie 20, 21500 Piikkiö, Finland
[10]Institut de Planétologie et d'Astrophysique de Grenoble, 414, Rue de la Piscine, Domaine Universitaire, 38400 St-Martin d'Hères, France
[11]Finnish Meteorological Institute, Observation services, Erik Palménin aukio 1, FI-00560 Helsinki, Finland
[12]ESA – ESTEC, Postbus 299, 2200AG Noordwijk, The Netherlands
[13]Institut für Geowissenschaften der Universität Heidelberg, Im Neuenheimer Feld 234-236, 69120 Heidelberg, Germany

*Corresponding author : Cecile.Engrand@csnsm.in2p3.fr

Manuscript pages: 19
Figures: 3
Tables: 4

Corresponding author:
Cecile Engrand
CSNSM CNRS/IN2P3-Univ. Paris Sud
Batiment 104
91405 Orsay Campus
Cecile.Engrand@csnsm.in2p3.fr
Tel : +33 1 69 15 52 95

## Abstract

COSIMA (COmetary Secondary Ion Mass Analyser) is a time-of-flight secondary ion mass spectrometer (TOF-SIMS) on board the Rosetta space mission. COSIMA has been designed to measure the composition of cometary dust grains. It has a mass resolution m/Δm of 1400 at mass 100 u, thus enabling the discrimination of inorganic mass peaks from organic ones in the mass spectra. We have evaluated the identification capabilities of the reference model of COSIMA for inorganic compounds using a suite of terrestrial minerals that are relevant for cometary science. Ground calibration demonstrated that the performances of the flight model were similar to that of the reference model. The list of minerals used in this study was chosen based on the mineralogy of meteorites, interplanetary dust particles and Stardust samples. It contains anhydrous and hydrous ferromagnesian silicates, refractory silicates and oxides (present in meteoritic Ca-Al-rich inclusions), carbonates, and Fe-Ni sulfides. From the analyses of these minerals, we have calculated relative sensitivity factors for a suite of major and minor elements in order to provide a basis for element quantification for the possible identification of major mineral classes present in the cometary grains.

## 1 Introduction

Comets spend most of their lifetime far away from the sun and are therefore only little affected by solar radiation. In addition, as they are small bodies, they are very likely not altered by internal differentiation. Therefore comets are considered to be among the most primitive objects in the Solar System and might even still contain residuals of the solar nebula. In other words, comets may have preserved refractory and/or volatile interstellar material left over from Solar System formation and can provide key information on the origin of our Solar System.

While remote observations allow measurements of collective properties of cometary dust, mass spectrometers flown on spacecraft allow the compositional analysis of individual particles. The latter technique was first introduced on the Giotto and Vega 1/2 missions to comet 1P/Halley (Kissel et al., 1986a; Kissel et al., 1986b). The measurements showed that in comet Halley's dust, a mineral component is mixed with organic matter in individual grains (Lawler and Brownlee, 1992).

Remote observations of comet C/1995 O1 (Hale-Bopp) and other bright comets, as well as laboratory analyses of cosmic dust of inferred cometary origin, showed that cometary dust is an unequilibrated, heterogeneous mixture of crystalline and glassy silicate minerals, organic refractory material, and other constituents such as iron sulfide and possibly minor amounts of iron oxides (Bradley, 2005; Crovisier et al., 1997; Dobrică et al., 2012; Hanner and Bradley, 2004, and references therein). The silicates are mostly Mg-rich, while Fe is distributed in silicates, sulfides, and FeNi metal. Remote infrared spectra of silicate emission features in comet Hale-Bopp have led to identification of the minerals forsterite and enstatite in both, amorphous and crystalline form. This mineralogy is consistent with the composition of chondritic porous anhydrous interplanetary dust particles (CP-IDPs) (e.g., Brunetto et al., 2011) and of UltraCarbonaceous Antarctic MicroMeteorites (UCAMMs) (Dobrică et al., 2012). The high D/H ratios of the organic refractory material in these IDPs (Messenger, 2002) and in UCAMMs (Duprat et al., 2010), as well as the physical and chemical structure of glassy silicate grains, suggest a primitive origin of cometary dust. Whether the components are of presolar origin is still a matter of debate. Carbon is enriched relative to CI chondrites; some of the C is in an organic phase (Jessberger et al., 1988).

"Ground truth" was provided by the Stardust mission which successfully returned in 2006 samples of dust collected in the coma of comet 81P/Wild 2 (Brownlee, 2014; Brownlee et al., 2006). The bulk of the Stardust samples appear to be weakly constructed mixtures of nanometer-sized grains, interspersed with much larger (>1 µm) ferromagnesian silicates, Fe-Ni sulfides, Fe-Ni metal, and others (Zolensky et al., 2006). The very wide variety of olivine and low-Ca pyroxene compositions in comet Wild 2 requires a wide range of formation conditions, probably reflecting very different formation locations in the protoplanetary disk (e.g., Frank et al., 2014). The restricted compositional ranges of Fe-Ni sulfides, the wide range for silicates, and the absence of hydrous phases indicate that comet Wild 2 likely experienced little or no aqueous alteration. Less abundant Wild 2 materials include refractory grains such as calcium-aluminum-rich inclusions (CAIs), high-temperature phases (Brownlee, 2014, and references therein), whose presence appears to require radial transport in the early protoplanetary disk.

Spitzer Space Telescope observations of comet 9P/Tempel 1 during the Deep Impact encounter revealed emission signatures that were assigned to amorphous and crystalline silicates, amorphous carbon, carbonates, phyllosilicates, polycyclic aromatic hydrocarbons, water gas and ice, and sulfides (Lisse et al., 2006). Good agreement is seen between the Tempel 1 ejecta spectra, the material emitted from comet Hale-Bopp, and the circumstellar material around the young stellar object HD100546 (Malfait et al., 1998). The atomic abundance of the observed material is consistent with solar and CI chondritic abundances. The presence of the observed mix of materials requires efficient methods of annealing amorphous silicates and mixing of high- and low-temperature phases over large distances in the early protosolar nebula.

In August 2014, the European Space Agency's spacecraft Rosetta arrived at Jupiter-family comet 67P/Churyumov-Gerasimenko (hereafter 67P/C-G). The Rosetta spacecraft carries eleven scientific instruments to study the nucleus of the comet as well as the gas, plasma, and particle environment in the inner coma as a function of heliocentric distance. On November 12, 2014, the lander spacecraft Philae has performed the first ever landing on a comet nucleus and provided *in situ* analysis of its physical and compositional properties (Gibney, 2014; Glassmeier et al., 2007, and references therein; Hand, 2014).

One of the core instruments of the Rosetta payload is the COmetary Secondary Ion Mass Analyser (COSIMA) that presently collects and analyzes the composition of dust grains in the coma of 67P/C-G (Kissel et al., 2009). COSIMA is a high-resolution time-of-flight secondary ion mass spectrometry (TOF-SIMS) instrument (Vickerman et al., 2013), which uses an indium primary ion beam to analyze the chemical composition of collected cometary grains. The mass resolution is m/Δm ∼ 1400 at 50% height (FWHM) of the peak at m/z=100 u. The bombardment of indium ions onto the sample produces secondary ions that are subsequently accelerated into a time-of-flight mass spectrometer, generating a secondary ion mass spectrum. By switching polarity of the mass spectrometer potentials, COSIMA is able to collect either positive or negative secondary ions. The goal of the COSIMA investigation is the *in situ* characterization of the elemental, molecular, mineralogical, and possibly isotopic composition of dust in the coma of comet 67P/C-G.

A twin of the COSIMA instrument flying on board Rosetta is located at the Max-Planck-Institut für Sonnensystemforschung (hereafter MPS) in Göttingen. This instrument serves as a reference instrument (Reference Model, RM) for the COSIMA flight instrument (named COSIMA XM). Pre-launch tests have shown that the performances of the RM and the XM are similar. Since the launch of Rosetta in 2004, the RM has been extensively used for laboratory calibration measurements. We have obtained a "library" of COSIMA mass spectra of well prepared and specially selected reference samples. Our reference samples are, among others, pure minerals expected to be present at the comet. These reference spectra will facilitate interpretation of the mass spectra expected from the comet with the COSIMA XM.

In this paper, we describe calibration measurements with the COSIMA RM that we performed with a set of mineral samples during recent years. A similar calibration campaign with samples of organic compounds is described in an accompanying paper (Le Roy et al., 2014).

## 2 Samples and methods

### 2.1 Sample selection and determination of compositions.

For our COSIMA reference measurements, we selected minerals that have either been detected in comets or that were identified in other primitive Solar System materials, namely meteorites (in particular carbonaceous chondrites) or interplanetary dust particles (IDPs) and Antarctic micrometeorites. The selected mineral groups include anhydrous silicates (in particular olivines, pyroxenes, and feldspars of different compositions), hydrated silicates, oxides and hydroxides, carbonates, sulfides, pure elements and alloys (Table 1). For the abundant minerals in comets, in particular anhydrous silicates, more than one sample was measured from the same mineral class (e.g., olivine). The samples were either purchased from a commercial provider (MPS samples - Krantz Mineral Shop in Bonn, Germany) or obtained from collections of the natural history museums in Los Angeles, London, Paris, and Vienna. A few samples were also provided through personal collaborations. The compositions of the mineral samples were either obtained from the literature, or were measured by electron microprobe at Univ. Paris VI, CAMPARIS. Major and minor elements were measured at 15 keV, 10 nA. Oxygen, carbon, and hydrogen were not measured but calculated by stoichiometry (for oxygen) or by difference (for carbon and hydrogen). The corresponding formula were calculated and compared to the theoretical values (Tables 1 and 2).

Table 1. Minerals analyzed with COSIMA RM including target types and numbers, and the sample preparation technique.

| Mineral family | Mineral Name | General Formula | Measured Formula | Provider (Origin)♮ | Target Type | Cosima Target Label | Preparation Technique |
|---|---|---|---|---|---|---|---|
| Ca-poor Px | Orthopyroxene | $(Mg,Fe)SiO_3$ | $(Mg_{0.9}Fe_{0.1})Si_{0.9}O_3$ | MPS – From M. Trieloff (Z31 Zabargad Island Kurat et al., 1993; Trieloff et al., 1997) | Au blank | 41E | Pressing |
| Ca-poor Px | Enstatite | $(Mg,Fe)SiO_3$ | $Mg_{0.9}SiO_3$ | CSNSM (MM, R2958, Bamle, Norway) | Ag blank | 49C (111) | Suspension |
| Ca-poor Px | Hypersthene | $(Mg,Fe)SiO_3$ | $(Mg_{0.7}Fe_{0.3})SiO_3$ | CSNSM - Los Angeles Museum | Ag blank | 496 (150) | Suspension |
| Ca-rich Px | Clinopyroxene | $CaMgSi_2O_6$ | $Ca_{0.7}Al_{0.1}Mg_{0.9}Fe_{0.1}Si_{1.8}O_6$ | MSP – From M. Trieloff (DW918 Witt-Eickschen et al., 2003)) | Au blank | 41D | Pressing |
| Ca-rich Px | Diopside | $CaMgSi_2O_6$ | $CaAl_{0.1}Mg_{0.9}Fe_{0.1}Si_2O_6$ | CSNSM-NHM Vienna (Madagaskar) | Ag blank | 49C (111) | Suspension |
| Ca-rich Px | Diopside | $CaMgSi_2O_6$ | $(Ca_{0.6}Na_{0.4})(Mg_{0.6}Al_{0.4})Si_2O_6$ | CSNSM-NHM London, BM 1906,382 (Italy) | Ag blank | 49D (114) | Suspension |
| Ca-rich Px | Augite | $(Ca,Na)(Mg,Fe,Al,Ti)(Si,Al)_2O_6$ | $(Ca_{0.9})(Mg_{0.8}Fe_{0.2})(Si_{1.8}Al_{0.2})O_6$ | CSNSM-NHM London (Daun tuff quarry, Germany) | Ag blank | 496 (150) | Suspension |
| Ca-rich Px | Hedenbergite | $CaFeSi_2O_6$ | $(Ca_{1.1})(Mg_{0.3}Fe_{0.5}Mn_{0.1})Si_2O_6$ | LPC2E-ISTO 90407 | Ag blank | 497 (136) | Suspension |
| Olivine | Forsterite | $Mg_2SiO_4$ | $Mg_2SiO_4$ | CSNSM – From A. Revcolevski (Synthetic mineral) | Ag blank | 496 (150) | Suspension |
| Olivine | Forsterite | $Mg_2SiO_4$ | $Mg_2SiO_4$ | CSNSM – From A. Revcolevski (Synthetic mineral) | Au blank | 420 | Pressing |
| Olivine | Olivine Zabargad | $(Mg,Fe)_2SiO_4$ | $(Mg_{1.8}Fe_{0.2})SiO_4$ | MPS – From M.Trieloff (Z104 Zabargad Island Kurat et al., 1993; Trieloff et al., 1997) | Au blank | 48C | Pressing |
| Olivine | Fayalite | $Fe_2SiO_4$ | $Fe_{1.9}SiO_4$ | CSNSM – From J. Borg | Ag blank | 496 (150) | Suspension |
| Feldspar | Albite | $NaAlSi_3O_8$ | $NaAlSi_3O_8$ | CSNSM (MM 118082, Ramona San Diego) | Ag blank | 496 (150) | Suspension |
| Feldspar | Anorthite | $CaAl_2Si_2O_8$ | $CaAl_2Si_2O_8$ | CSNSM-NHM Vienna, (T. de la Foya, Austria) | Ag blank | 497 (136) | Suspension |
| Feldspar | Plagioclase | $(Na,Ca)(Si,Al)_4O_8$ | $(Na_{0.5}Ca_{0.5})(Si_{2.5}Al_{1.5})O_8$ | CSNSM-NHM Vienna (Tanzmeister, Austria) | Ag blank | 497 (136) | Suspension |
| Feldspar | Plagioclase | $(Na,Ca)(Si,Al)_4O_8$ | $(Na_{0.5}Ca_{0.5})(Si_{2.5}Al_{1.5})O_8$ | CSNSM-NHM Vienna (Tanzmeister, Austria) | Ag blank | 48B (AG57) | Pressing |
| Feldspar | Orthoclase | $KAlSi_3O_8$ | $(Na_{0.3}K_{0.6})AlSi_3O_8$ | CSNSM-NHM London (Moon Stone, Sri Lanka) | Ag blank | 496 (150) | Suspension |
| Feldspathoid | Nepheline | $(Na,K)AlSiO_4$ | $(Na_{0.6}Ca_{0.3})AlSiO_4$ | CSNSM-NHM London (York River, Ontario CA) | Ag blank | 497 (136) | Suspension |
| Hydr. silicate | Fuchsite | $KAl_2(Si_3Al)O_{10}(OH,F)_2$ | $Na_{0.1}K_{0.5}Si_{3.2}Al_{2.8}Fe_{0.1}O_{10}(OH)_{1.8}*$ | CSNSM | Ag blank | 49D (114) | Suspension |
| Hydr. silicate | Richterite | $Na(CaNa)(Mg,Fe)_5[Si_8O_{22}](OH)_2$ | $Na_{0.9}Al_{0.3}K_{0.2}Ca_{1.6}(Mg_{4.6}Fe_{0.4})[Si_8O_{21.2}](OH)_{4.8}*$ | MPS (Bancroft Ontario, Canada) | Ag blank | 4B0 (147) | Suspension |
| Hydr. silicate | Smectite | $Ca_{0.25}(Mg,Fe)_3((Si,Al)_4O_{10})(OH)_2 \cdot nH_2O$ | $Ca_{0.2}(Mg_{0.1}Fe_{2.5})((Si_4Al_{0.1})O_{10}(OH)_2 \cdot 2H_2O*$ | CSNSM (Bowling, Le Lamentin, Martinique | Au blank | 422 | Pressing |
| Hydr. silicate | Talc | $Mg_3Si_4O_{10}(OH)_2$ | $Mg_{3.4}Si_{3.8}O_{10}(OH)_2 \cdot 4H_2O*$ | CSNSM-Museum Lauzenac Ariege | Ag blank | 49C (111) | Suspension |
| Carbonate | Dolomite | $CaMg(CO_3)_2$ | $Ca(Mg_{0.8}Fe_{0.2})(CO_3)_2^{\#}$ | MPS (Vegarsheien, Norway) | Ag blank | 4AF (142) | Suspension |
| Carbonate | Calcite | $CaCO_3$ | $Ca_{1.1}CO_3^{\#}$ | MPS (Creel Chihuahua, Mexico) | Ag blank | 4AF (142) | Suspension |
| Melilite | Melilite | $(Ca,Na)_2(Al,Mg,Fe)(Si,Al)_2O_7$ | $(Ca_{1.8}Na_{0.1})(Al_{0.6}Mg_{0.3}Fe_{0.1})(Si_{1.6}Al_{0.4})O_7$ | CSNSM-MNHN Paris (Vesuvius) | Ag blank | 498 (143) | Suspension |
| Melilite | Åkermanite | $Ca_2Mg[Si_2O_7]$ | $Ca_2Mg[Si_2O_7]$ | CSNSM-Dr.Morioka Japan (Synthetic mineral) | Ag blank | 496 (150) | Suspension |
| Oxide | Ilmenite | $FeTiO_3$ | $(Fe_{0.8}Mg_{0.2})TiO3$ | MPS (Flekkefjord, Norway) | Ag blank | 4B0 (147) | Suspension |
| Oxide | Magnetite | $Fe_3O_4$ | $Fe_{2.5}O_4$ (O measured as FeO) | MPS (Minas Gerais, Brasil) | Ag blank | 4B0 (147) | Suspension |
| Oxide | Corundum | $Al_2O_3$ | $Al_2O_3$ | CSNSM-NHM Vienna (Ceylon) | Ag blank | 4B0 (147) | Suspension |
| Sulfide | Sphalerite | $[(Zn, Fe)S]$ | $ZnS$ | CSNSM (Picos de Europa, Spain) | Ag blank | 4AF (142) | Suspension |
| Sulfide | Pyrite | $FeS_2$ | $FeS_{2.0}$ | CSNSM-CRPG | Au blank | 421 | Pressing |
| Sulfide | Pentlandite | $(Fe,Ni)_9S_8$ | $(Fe_{4.4}Ni_{4.8}Co_{0.1})S_8$ | CSNSM-CRPG | Ag blank | 49D (114) | Suspension |
| Sulfide | Pyrrhotite | $Fe_{(1-x)}S$ (x = 0 - 0.17) | $FeS$ | CSNSM-CRPG | Ag blank | 49D (114) | Suspension |
| --- | Substrate gold | ---------- | -------- | -------- | Au blank | 41D/41E | No sample |
| --- | Substrate silver | ---------- | -------- | -------- | Ag blank | 49C (111) | No sample |

*H calculated by difference
#C calculated by difference
♮Provider of the minerals, and sampling location, when available. MPS : Max Planck Institut für Sonnensystemforschung (Göttingen, Germany). CSNSM : Centre de Sciences Nucléaires et de Sciences de la Matière (Orsay France). LPC2E : Laboratoire de Physique et Chimie de l'Environnement et de l'Espace (Orléans France). ISTO : Institut des Sciences de la Terre d'Orléans (France). NHM : Natural History Museum (Vienna Austria, or London UK), MNHN : Museum National d'Histoire Naturelle (Paris France). CRPG : Centre de Recherches Pétrographiques et Géochimiques (Nancy France).

Table 2. Composition of the minerals measured by electron microprobe (atomic percent).

| Mineral Family | Mineral Name | O | Na | Mg | Al | Si | P | S | K | Ca | Ti | V | Cr | Mn | Fe | Co | Ni | Cu | Zn | H* | C# | Total |
|---|---|---|---|---|---|---|---|---|---|---|---|---|---|---|---|---|---|---|---|---|---|---|
| Ca-poor Px | Orthopyroxene | 60.77 | b.d. | 17.41 | 0.80 | 19.02 | - | - | - | 0.14 | b.d. | - | 0.04 | 0.04 | 1.75 | - | 0.04 | - | - | | | 100.00 |
| Ca-poor Px | Enstatite | 60.13 | b.d. | 18.94 | 0.07 | 20.15 | b.d. | b.d. | b.d. | b.d. | b.d. | - | b.d. | b.d. | 0.71 | - | b.d. | - | - | | | 100.00 |
| Ca-poor Px | Hypersthene | 59.94 | b.d. | 13.67 | 0.90 | 19.34 | b.d. | b.d. | b.d. | 0.56 | 0.05 | - | 0.05 | 0.12 | 5.36 | - | b.d. | - | - | | | 100.00 |
| Ca-rich Px | Clinopyroxene | 61.91 | 0.30 | 9.24 | 1.08 | 18.60 | - | b.d. | - | 7.20 | 0.16 | - | 0.09 | 0.03 | 1.35 | - | 0.03 | - | - | | | 100.00 |
| Ca-rich Px | Diopside. (Madagaskar) | 60.00 | 0.20 | 9.05 | 1.01 | 19.52 | b.d. | b.d. | b.d. | 9.55 | b.d. | - | b.d. | 0.04 | 0.63 | - | b.d. | - | - | | | 100.00 |
| Ca-rich Px | Diopside (Italy) | 59.95 | 4.04 | 5.90 | 3.66 | 19.95 | 0.07 | b.d. | b.d. | 5.95 | 0.03 | - | b.d. | 0.36 | 0.11 | - | b.d. | - | - | | | 100.00 |
| Ca-rich Px | Augite | 59.86 | 0.39 | 7.58 | 2.36 | 18.28 | b.d. | b.d. | b.d. | 9.12 | 0.41 | - | 0.08 | 0.04 | 1.88 | - | b.d. | - | - | | | 100.00 |
| Ca-rich Px | Hedenbergite | 60.08 | - | 3.21 | - | 20.15 | - | - | - | 10.74 | - | - | - | 0.52 | 5.30 | - | - | - | - | | | 100.00 |
| Olivine | Synthetic Forsterite | 57.21 | b.d. | 28.36 | b.d. | 14.43 | b.d. | b.d. | b.d. | b.d. | b.d. | - | b.d. | b.d. | b.d. | - | b.d. | - | - | | | 100.00 |
| Olivine | San Carlos Olivine | 56.54 | - | 25.82 | - | 14.81 | - | - | - | b.d. | - | - | b.d. | - | 2.73 | - | 0.10 | - | - | | | 100.00 |
| Olivine | Fayalite | 57.28 | 0.07 | 0.18 | 0.12 | 14.46 | b.d. | b.d. | b.d. | b.d. | b.d. | - | b.d. | b.d. | 27.89 | - | b.d. | - | - | | | 100.00 |
| Feldspar | Albite | 61.54 | 7.57 | b.d. | 7.85 | 22.95 | b.d. | b.d. | 0.04 | 0.05 | b.d. | - | b.d. | b.d. | b.d. | - | b.d. | - | - | | | 100.00 |
| Feldspar | Anorthite | 61.55 | b.d. | b.d. | 15.29 | 15.40 | b.d. | b.d. | b.d. | 7.76 | b.d. | - | b.d. | b.d. | b.d. | - | b.d. | - | - | | | 100.00 |
| Feldspar | Plagioclase | 61.49 | 3.56 | b.d. | 11.64 | 19.02 | b.d. | b.d. | 0.20 | 4.03 | b.d. | - | b.d. | b.d. | 0.07 | - | b.d. | - | - | | | 100.00 |
| Feldspar | Orthoclase | 61.62 | 2.67 | b.d. | 7.82 | 23.01 | b.d. | b.d. | 4.80 | 0.09 | b.d. | - | b.d. | b.d. | b.d. | - | b.d. | - | - | | | 100.00 |
| Feldspathoid | Nepheline | 58.45 | 9.13 | b.d. | 14.33 | 14.31 | b.d. | b.d. | b.d. | 3.78 | b.d. | - | b.d. | b.d. | b.d. | - | b.d. | - | - | | | 100.00 |
| Hydrated silicate | Fuchsite | 58.51 | 0.32 | 0.20 | 13.82 | 15.40 | b.d. | b.d. | 2.25 | b.d. | 0.32 | - | 0.05 | b.d. | 0.39 | - | b.d. | - | - | 8.7* | | 100.00 |
| Hydrated silicate | Richterite | 55.64 | 1.90 | 9.90 | 0.58 | 17.17 | b.d. | b.d. | 0.38 | 3.37 | b.d. | - | b.d. | 0.03 | 0.82 | - | b.d. | - | - | 10.2* | | 100.00 |
| Hydrated silicate | Smectite | 52.13 | 0.07 | 0.43 | 0.55 | 14.92 | b.d. | b.d. | 0.15 | 0.64 | b.d. | - | b.d. | b.d. | 9.34 | - | 0.03 | - | - | 21.7* | | 100.00 |
| Hydrated silicate | Talc | 48.11 | b.d. | 10.27 | b.d. | 11.32 | b.d. | b.d. | b.d. | b.d. | b.d. | - | b.d. | b.d. | 0.09 | - | b.d. | - | - | 30.2* | | 100.00 |
| Carbonate | Dolomite | 59.99 | b.d. | 8.37 | b.d. | b.d. | b.d. | b.d. | b.d. | 10.01 | b.d. | b.d. | b.d. | 0.05 | 1.61 | - | - | - | - | | 20.0# | 100.00 |
| Carbonate | Calcite | 59.56 | b.d. | b.d. | b.d. | b.d. | b.d. | b.d. | b.d. | 21.34 | b.d. | - | b.d. | b.d. | b.d. | b.d. | - | - | - | | 19.1# | 100.00 |
| Melilite | Melilite | 58.39 | 1.03 | 2.80 | 8.12 | 13.28 | b.d. | b.d. | 0.11 | 15.41 | b.d. | - | b.d. | b.d. | 0.85 | - | b.d. | - | - | | | 100.00 |
| Melilite | Åkermanite | 58.39 | b.d. | 8.28 | b.d. | 16.79 | b.d. | b.d. | b.d. | 16.54 | b.d. | - | b.d. | b.d. | b.d. | - | b.d. | - | - | | | 100.00 |
| Oxide | Ilmenite | 59.66 | b.d. | 4.05 | b.d. | b.d. | b.d. | b.d. | b.d. | b.d. | 19.28 | - | 0.05 | 0.11 | 16.81 | - | 0.04 | - | - | | | 100.00 |
| Oxide | Magnetite | 61.29 | b.d. | 0.08 | 0.23 | 0.05 | b.d. | b.d. | b.d. | b.d. | b.d. | - | b.d. | 0.02 | 38.33 | - | b.d. | - | - | | | 100.00 |
| Oxide | Corundum | 60.01 | - | - | 39.99 | b.d. | b.d. | b.d. | b.d. | b.d. | b.d. | - | b.d. | b.d. | b.d. | - | b.d. | - | - | | | 100.00 |
| Sulfide | Sphalerite | - | - | - | - | - | - | 50.71 | - | - | - | - | - | - | 0.05 | - | - | - | 49.24 | | | 100.00 |
| Sulfide | Pyrite | - | - | - | - | - | - | 65.99 | - | - | - | - | b.d. | 0.05 | 33.97 | b.d. | b.d. | b.d. | - | | | 100.00 |
| Sulfide | Pentlandite | - | - | - | - | - | - | 46.30 | - | - | - | - | b.d. | b.d. | 25.70 | 0.35 | 27.65 | b.d. | b.d. | | | 100.00 |
| Sulfide | Pyrrhotite | - | - | - | - | - | - | 49.63 | - | - | - | 0.14 | b.d. | b.d. | 50.21 | b.d. | 0.03 | b.d. | - | | | 100.00 |

*H calculated by difference
#C calculated by difference
b.d. : below detection limit

## 2.2 COSIMA substrates and sample preparation

The COSIMA XM is equipped with a set of 72 substrates of different types to collect dust grains in the cometary coma (Genzer and Rynö, 2010):

_ 34x Gold black
_ 12x Silver black
_ 16x Platinum black
_ 3x Palladium black
_ 7x Silver blank

The metallic black were obtained by deposition of nanometer-sized grains, formed by condensation from the vapor phase in the case of Au and Ag, or by an electrochemical procedure in the case of Pt and Pd (Hornung et al., 2014). To ease sample preparation for the laboratory calibration campaign, we selected blank silver substrates in most cases and in a few cases blank gold substrates (Table 1).

The blank silver targets were cleaned in an ultrasonic bath for 15 min, first in isopropanol and then in distilled water. The gold blank targets were cleaned with acetone, and heated to 900°C for 5 min. The heating of the target removes any organics from the target and softens the gold.

The samples procured by MPS (see Table 1) were crushed to pieces of several millimeters to one centimeter in size with a jaw crusher. They were then cleaned for 10 min in ethanol followed by 10 min in deionized water using an ultrasonic bath. After inspection by eye and optical microscope to identify pure and clean specimen of the respective mineral, a few selected specimens were then ground in an electrical ball mill down to a smallest grain size of approximately 25 µm. The powder obtained was then sieved with a set of stainless steel sieves with mesh sizes between 25 µm and 200 µm. The fraction with grain size between 25 µm and 50 µm was usually selected for COSIMA measurements. Samples from MPS (Zabargad olivine and clino- and orthopyroxene; Kurat et al., 1993; Trieloff et al., 1997; Witt-Eickschen et al., 2003) were prepared by standard mineral separation techniques, i.e., several cycles of hand-picking of coarse grained material, crushing, sieving, washing, and occasionally magnetic separation. Samples from CSNSM were washed in acetone, then crushed in an agate mortar and visually examined in order to pick a fragment relevant to the mineral type.

Two different types of sample preparation were used (Table 1): suspension (at MPS) or pressing in gold foil (at CSNSM). In the first case, the sieved minerals was suspended in water and applied to the metal target with a pipette in the field of view of a laboratory microscope. Approximately 1 µl was deposited onto the substrate with a pipette, creating a 1 mm droplet on the surface of the target. After evaporation of the solvent, a homogeneous surface coverage was usually visible on the substrate. For the pressing method, small fragments (~ 50 µm) of the minerals were selected using a binocular and pressed into the blank gold substrates using a microcrusher at CSNSM, which is usually used to press micrometeorites or IDPs into gold foils. In the microcrusher, the sample was crushed into the foil with a disk of fused silica previously cleaned for 10 min with ethanol followed by 10 min in deionized water using an ultrasonic bath.

### 2.3 Measurement strategy with the COSIMA RM and selection of best mass spectra

Positive and negative secondary ion mass spectra have been obtained for the 31 selected minerals with the COSIMA RM at MPS. Each sample was measured at least at two locations as a set of 4x4 or 5x5 raster with a separation of the raster points of 50 µm. The measurement time per spectrum was typically 5 min. Background spectra were systematically acquired outside of the sample (see below).

In order to establish the intensities and intensity ratios for several relevant elements for each mineral analyzed, three to six positive mass spectra were selected for evaluation. The same coordinates on the target were used for selection of positive as well as negative spectra. The mass spectra were selected based on the following criteria: the spectra had to be on the mineral grains and the level of contamination in the spectra had to be as low as possible. Common contaminants found in TOF-SIMS analyses include (i) organic compounds such as polydimethylsiloxane (PDMS - a silicone oil) and phthalates, and (ii) ions such as $Na^+$, $K^+$, $Cl^-$, $SO_2^-$, $SO_3^-$, and $HSO_4^-$. Mass spectra measured outside of the minerals on the substrate were used as contamination control. Features related to common contaminants for these analyses will be described in Hilchenbach et al. (in prep.), therefore this topic will not be discussed further. Last but not least, particular attention was paid to the characteristic peaks of each mineral: presence of individual elements and correlated elements, as well as their corresponding intensities.

### 2.4 Raman measurements

Raman spectra of all mineral samples were measured using a laboratory Raman spectrometer (model alpha300 R, WITec, Ulm, Germany) in order to verify the mineral identification. The confocal Raman spectrometer is equipped with a polarized fiber optic coupled 532 nm laser. Raman scattered light and fluorescence emission is transmitted through a beamsplitter, a laser notch filter, a long wavelength filter, and a 50 µm optical fiber to a spectrometer with a Peltier-cooled CCD detector. The wavenumber range of the spectrometer is 150 $cm^{-1}$ to 3800 $cm^{-1}$ with 5 $cm^{-1}$ spectral resolution. The microscopic system accommodates three objectives with increasing magnifying power and numeric aperture (10x, 50x, and 100x). Metal targets with mineral grains were placed on the piezo-driven x-y scan table beneath the objective coupled to the z-axis focusing unit. The coarse sample grain surfaces were monitored with a CCD video camera prior to and following the Raman scans with diffraction limited optics. The excitation intensity of the laser system was adjusted prior to the depth scan with a variable slit between the laser and the transmitting optical fiber to maximize the recorded Raman and fluorescence emission while keeping sample alterations due to heating at a minimum. Since the sample composition and therefore absorption and refractive index were not spatially uniform, sample spot deterioration could not be ruled out prior to matrix scans for the whole area. The excitation intensity varies from 0.4 mW to 5 mW for samples with high and low absorption. The effective measurement time interval and laser illumination was 0.2 s for each scan matrix point. For each sample, the spectral data was obtained from two areas 80 x 80 µm² that had been previously analyzed by COSIMA. Two kinds of Raman analyses were made: first a slice cutting through a selected area was scanned in depth mode along a line parallel to the x-axis in the x-z-plane, and then, an image mode scan was made parallel to the focal plane in an x-y plane encompassing the sample surface. In the latter case, fractions of the rectangular scan matrix were in or out of focus due to the surface roughness of the mineral grain samples. The recorded fluorescence and Raman emission spectra were corrected for cosmic ray particle events. Spectra were summed up and averaged for selected adjoining measurement points within each scan to improve the signal-to-noise ratio. False color images for both scan modes were plotted for selected spectral bandwidth, resulting in depth and image scans each representing other spectral features and thereby allowing the spatial identification and feasible separation of the emission sources. The minerals were identified by comparison of the observed Raman scattered lines with a database of Raman spectra of minerals accessible via the RRUFF Project webpage (RRUFF Project).

### 2.5 Sputtering

To clean the mineral surface before analysis, sputtering with a direct current or long pulses is often used in SIMS (Stephan, 2001). During sputtering, the mineral surface is exposed to a much higher ion dose than during analysis when short pulses are used. This increased ion dose efficiently removes any organic molecules and other undesired components covering the mineral surface, thus increasing the ion signals obtained from the mineral. In addition to removing any contaminants, the sputtering also makes the mineral matrix more homogenous and causes amorphization (e.g., Stephan, 2001), which means a more stable ion signal can be obtained during subsequent analyses. The sputtering time needed to obtain these effects depends on the ion beam and current used, and size of area sputtered, however a couple of minutes of sputtering is usually sufficient (Siljeström et al., 2010; Stephan, 2001). Most studies of sputtering of minerals have been done on flat mineral surfaces with either $Ar^+$, $C_{60}^+$, $Bi^+$, $Cs^+$, and $O^{+/-}$, which are the sputter ion beams most frequently found on commercial TOF-SIMS instruments (Siljeström et al., 2011; Stephan, 2001). So far, no studies on the sputtering of mineral grains with an indium primary beam have been performed. Therefore, studies on the effects of sputtering of mineral grains had first to be performed in the RM before it can be used on samples collected in space. The sputtering experiments were executed according to the following protocol: 5 min sputtering followed by analysis during 5 min, and this sequence was repeated up to 10 times. The primary emission current for sputtering was 10 µA (continuous beam), and 5 µA for the analysis beam (pulsed beam). We will not further discuss the sputtering experiments on mineral grains performed in the RM, as no significant effect on the stability of the secondary ion beam was demonstrated. Sputtering was, however, useful for cleaning the sample surface from PDMS contamination.

### 3 Results and discussion

Measurements of all mineral samples listed in Tables 1 and 2 were performed with the COSIMA RM instrument at MPS. Representative spectra of a selection of minerals are displayed in Figure 1. Most minerals have a signature in positive secondary ions (Fig. 1a to 1e), and Fe-sulfides show the sulfur signature in negative secondary ion spectra (Fig. 1f). Figure 2 shows details of some elemental peaks presenting the discrimination between the inorganic peak and the organic peak present at each mass.

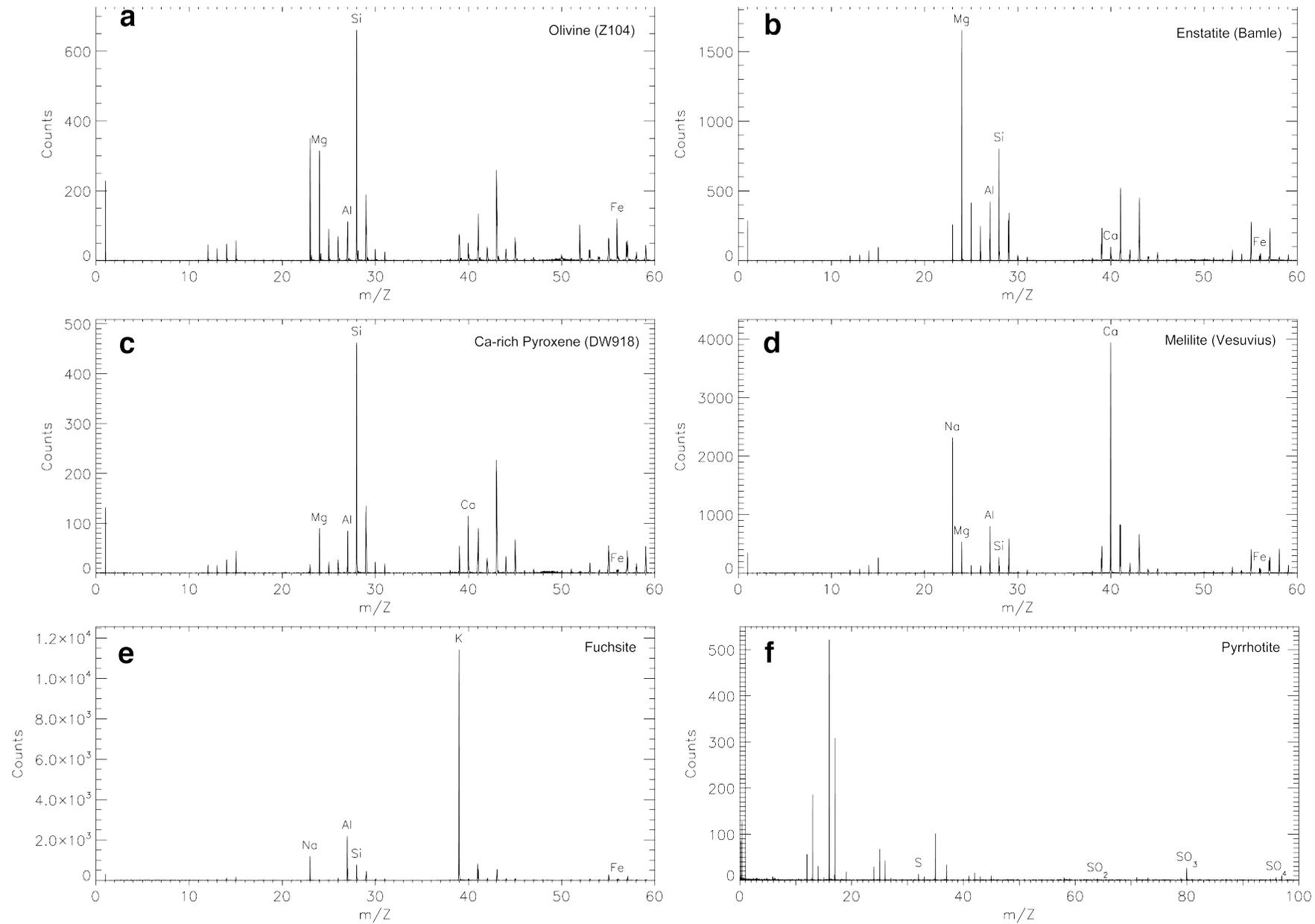

Figure 1 : Representative spectra of some minerals analyzed in this study (a-e: positive secondary ions, f: negative secondary ions). a) Mg-rich olivine (Z104); b) Mg-rich Ca-poor pyroxene (Enstatite Bamle); c) Ca-rich pyroxene (DW918); d) Melilite (Vesuvius); e) Hydrated mineral fuchsite; f) Fe-sulfide pyrrhotite. See Tables 1 and 2 for description and compositions of the minerals.

### 3.1 Relative sensitivity factors

In order to quantify SIMS results, to calculate element ratios from secondary ion ratios, relative sensitivity factors (RSFs) are needed. Positive spectra were used for the quantification of the RSFs. Negative spectra are used to demonstrate the presence of S-rich compounds, especially in the case of Fe-sulfides. As no normalizing element is present in negative spectra for Fe sulfides, no RSF could be calculated for sulfur.

For a known element atomic ratio $E/E_0$, the RSF can be calculated from the secondary ion intensities $SI$:

$$RSF(E, E_0) = \frac{SI(E)/SI(E_0)}{E/E_0}$$

The elemental $E/E_0$ ratio is known from the composition of the mineral, and the normalizing element $E_0$ is usually one of the most abundant specie (i.e., Si, Mg, or Fe). For silicate minerals, such RSFs are usually reported relative to Si since it is the only element besides O being present in all silicates. Oxygen, on the other hand, has a very low ionization probability for positive secondary ions, which makes it not suitable as a reference element. RSFs are obtained by analyzing standard materials with known elemental composition under the same conditions as the samples to be analyzed (i.e., cometary grains in this case). Following data evaluation steps as described by Stephan (2001), secondary ion ratios for numerous elements relative to Si were obtained. For some mineral samples, either no statistically significant Si element data were available or the Si peaks were compromised by silicone oil contamination. In such cases, we used Mg as a reference element and renormalized the result to Si using an RSF(Mg/Si) of 3.71 (geometric mean of RSF(Mg/Si) values calculated for minerals with reliable Si data). If Si and Mg normalization failed, Fe was used as a reference element, and an RSF(Fe/Si) of 1.81 (geometric mean of RSF(Fe/Si) values calculated for minerals with reliable Si data) was applied. Table 3 shows all results that were used to calculate (geometric) mean values for the RSFs. Figure 3 shows a comparison between RSFs calculated in this study with data from Stephan (2001) for a commercial TOF-SIMS instrument that uses a 25 keV $^{69}$Ga$^+$ primary ion beam. The general trend for both primary ion species is the same, and ionization probability mainly depends on the ionization energy of a given element. For some elements, only limited data are available, e.g., V, Co, and Ni were only measured reliably in one mineral each. This might explain why the RSFs for Co and Ni do not seem to follow the general trend. Therefore, we recommend for these elements, RSFs that have been calculated from Fe-normalized values from Stephan (2001). Table 4 presents mean RSFs normalized to Si, Mg, and Fe.

Table 3. Relative sensitivity factors for positive secondary ions normalized to Si obtained with the COSIMA RM instrument at MPS.

| Sample | O | Na | Mg | Al | Si | K | Ca | Ti | V | Cr | Mn | Fe | Co | Ni |
|---|---|---|---|---|---|---|---|---|---|---|---|---|---|---|
| Enstatite Bamle | 0.00060(16) | — | 2.56(3) | — | ≡1 | — | — | — | — | — | — | 1.05(18) | — | — |
| Hypersthene | 0.0010(3) | — | 2.01(5) | — | ≡1 | — | 12.3(4) | 4.6(9) | — | — | 3.6(5) | 1.67(5) | — | — |
| Clinopyroxene | 0.0027(8) | 24.2(10) | ≡3.71 | — | — | — | 5.24(13) | 1.7(8) | — | 1.8(6) | — | 1.74(13) | — | — |
| Diopside Madagascar | 0.0009(3) | — | 3.06(5) | — | ≡1 | — | 7.84(11) | — | — | — | — | 3.3(3) | — | — |
| Diopside San Marcel | 0.0025(5) | 171(6) | 6.3(2) | 9.0(7) | ≡1 | — | 15.3(6) | — | — | — | — | — | — | — |
| Augite | 0.00061(19) | 50.1(9) | 3.77(7) | — | ≡1 | — | 5.60(9) | 4.8(2) | — | — | — | 1.8(5) | — | — |
| Hedenbergite | 0.00057(17) | — | 3.46(8) | — | ≡1 | — | — | — | — | — | 3.0(2) | 1.87(4) | — | — |
| Olivine Zabargad | 0.00039(11) | — | 3.49(3) | — | ≡1 | — | — | — | — | — | — | 2.17(3) | — | — |
| Fayalite | 0.0011(3) | — | 6.3(4) | — | ≡1 | — | — | — | — | — | — | 1.07(3) | — | — |
| Albite | 0.00039(13) | 18.6(3) | — | 4.25(15) | ≡1 | — | — | — | — | — | — | — | — | — |
| Anorthite | 0.00042(18) | — | — | 3.27(8) | ≡1 | — | 5.49(11) | — | — | — | — | — | — | — |
| Plagioclase (497) | 0.0008(2) | 8.28(12) | — | 1.45(6) | ≡1 | 62(2) | 4.69(8) | — | — | — | — | — | — | — |
| Plagioclase (48B) | 0.00083(14) | 18.3(2) | — | 4.08(11) | ≡1 | — | — | — | — | — | — | — | — | — |
| Orthoclase | 0.0009(4) | 10.9(4) | — | 3.4(3) | ≡1 | 17.6(7) | — | — | — | — | — | — | — | — |
| Nepheline | 0.0008(2) | 10.42(15) | — | 2.07(5) | ≡1 | — | 5.13(9) | — | — | — | — | — | — | — |
| Fuchsite | 0.0010(2) | 71.7(14) | 4.8(2) | 3.03(9) | ≡1 | 96.8(19) | — | 2.38(14) | — | 3.9(4) | — | 1.9(3) | — | — |
| Richterite | 0.0012(3) | 57.7(9) | 3.61(6) | — | ≡1 | 75.0(15) | 13.9(2) | — | — | — | — | 3.34(13) | — | — |
| Smectite (422) | 0.00030(5) | 127.0(11) | 3.37(6) | 4.9(5) | ≡1 | — | 9.72(10) | — | — | — | — | 0.932(9) | — | — |
| Smectite (49C) | 0.00033(8) | — | 5.32(11) | 9.4(4) | ≡1 | 75.7(18) | 9.14(14) | — | — | — | — | 1.57(2) | — | 2.4(11) |
| Talc | 0.00067(11) | — | 1.806(19) | — | ≡1 | — | — | — | — | — | — | — | — | — |
| Dolomite | 0.0010(3) | — | ≡3.71 | — | — | — | 7.59(11) | — | — | — | 2.3(7) | 1.0(3) | — | — |
| Melilite | 0.0013(3) | 115(2) | 9.5(2) | 1.14(13) | ≡1 | 118(5) | 10.9(2) | — | — | — | — | 5.04(18) | — | — |
| Akermanite | 0.00085(14) | — | 2.29(4) | — | ≡1 | — | 3.51(5) | — | — | — | — | — | — | — |
| Ilmenite | 0.0012(5) | — | ≡3.71 | — | — | — | — | — | — | — | — | 1.88(5) | — | — |
| Pentlandite | — | — | — | — | — | — | — | — | — | — | — | ≡1.81 | 0.73(8) | — |
| Pyrrhotite | — | — | — | — | — | — | — | — | 3.5(5) | — | — | ≡1.81 | — | — |

Errors referring to the last significant digits are given in parentheses (i.e. RSF(Fe/Si) for Enstatite Bamle = 1.05 ± 0.18). *n* gives the number of samples measured to calculate the geometric mean values. Two samples on different substrates were analyzed for plagioclase and smectite, respectively. For Co and Ni, recommended values given in italics are derived from literature values (Stephan, 2001).
*Numbers for O are calculated from positive spectra, hence the low values, comparable to those of Stephan (2001).

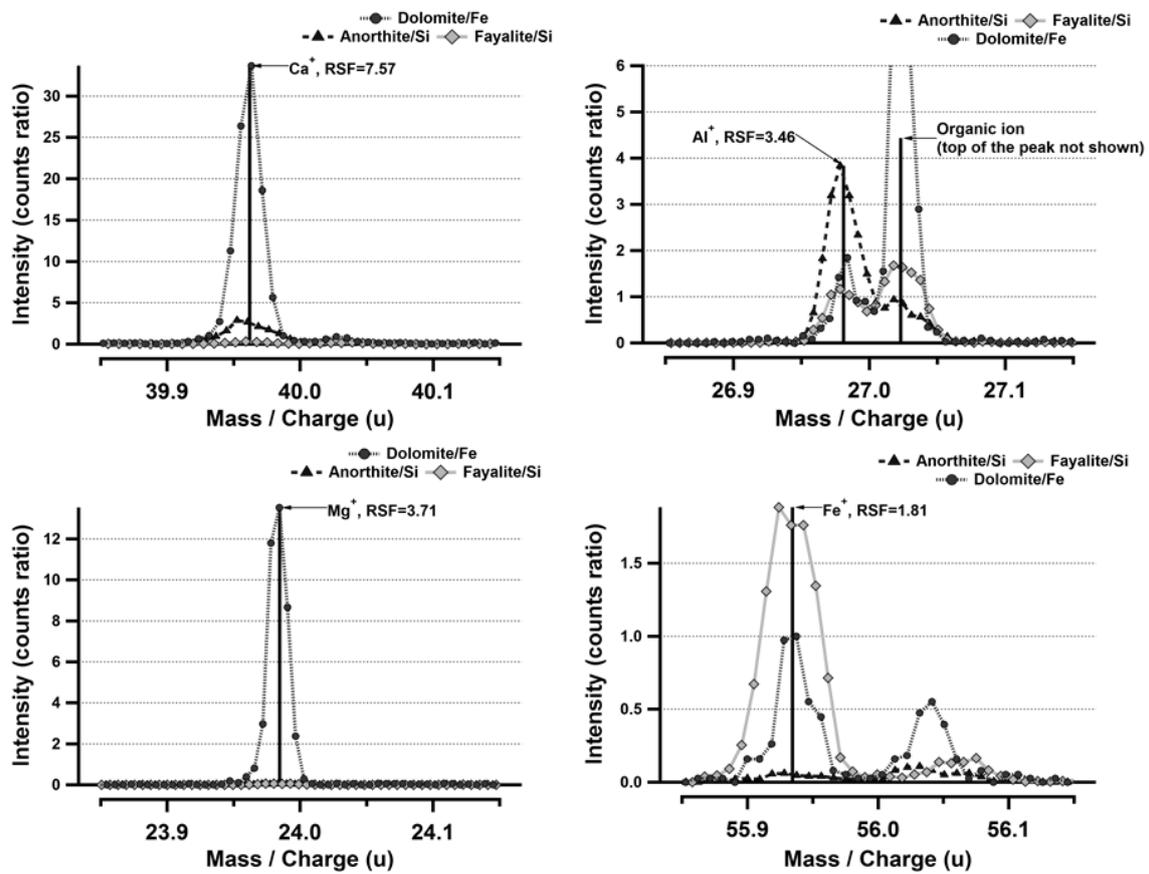

Figure 2: Details of Mg$^+$, Al$^+$, Ca$^+$, and Fe$^+$ peaks showing the discrimination between the inorganic (to the left of the integer mass/charge) and organic peaks (to the right of the integer mass/charge), thus allowing the quantification of relative sensitivity factors, normalized to Si, Mg or Fe.

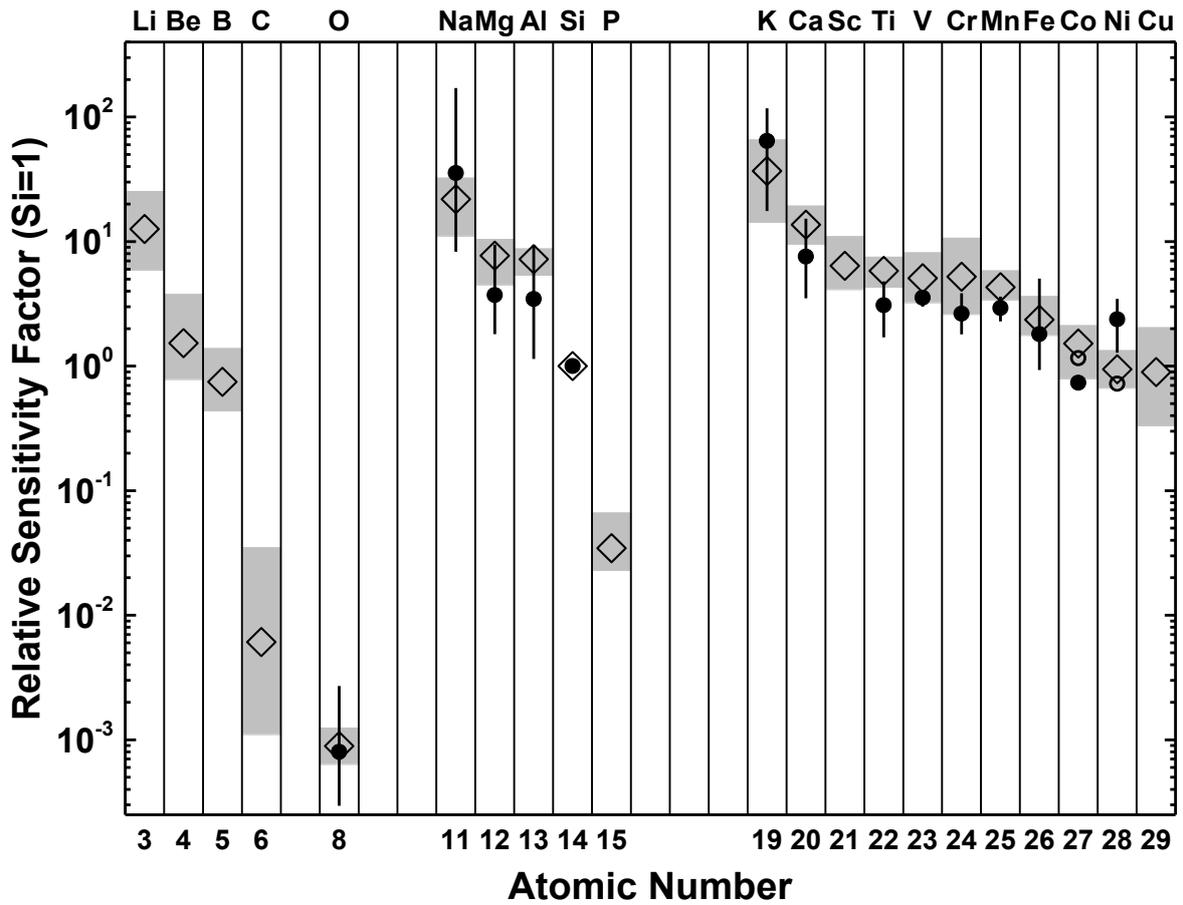

Figure 3: Mean RSFs normalized to Si for major and minor elements versus atomic number obtained from positive ion spectra of various mineral samples. Filled circles are geometric mean values from Table 3. The vertical bars show the range of individual values obtained for different minerals, except for V, Co, and Ni for which they represent statistical errors. Open circles for Co and Ni are derived from literature values (Stephan, 2001). For comparison, RSFs from Stephan (2001) for a commercial TOF-SIMS instrument with a 25 keV $^{69}$Ga$^+$ primary ion beam and that were obtained from a suite of glass standards are shown as open diamonds. For these, the variation range is shown in gray.

Table 4. Recommended COSIMA relative sensitivity factors for positive secondary ions normalized to Si, Mg, and Fe.

| Element | RSF (Si≡1) | $n$ | RSF (Mg≡1) | $n$ | RSF (Fe≡1) | $n$ |
|---|---|---|---|---|---|---|
| O  | $0.00080^{+0.00060}_{-0.00034}$ | 24 | $0.00022^{+0.00017}_{-0.00010}$ | 24 | $0.00044^{+0.00035}_{-0.00019}$ | 24 |
| Na | $35^{+67}_{-23}$ | 12 | $8.3^{+12.7}_{-5.0}$ | 12 | $18^{+35}_{-12}$ | 12 |
| Mg | $3.7^{+2.2}_{-1.4}$ | 15 | ≡1 | 18 | $2.0^{+1.4}_{-0.8}$ | 18 |
| Al | $3.5^{+3.2}_{-1.7}$ | 11 | $0.78^{+0.89}_{-0.42}$ | 11 | $1.8^{+2.8}_{-1.1}$ | 11 |
| Si | ≡1 | 21 | $0.27^{+0.16}_{-0.10}$ | 15 | $0.53^{+0.35}_{-0.21}$ | 12 |
| K  | $64^{+62}_{-31}$ | 6 | $13^{+10}_{-6}$ | 6 | $27^{+24}_{-13}$ | 6 |
| Ca | $7.6^{+4.3}_{-2.7}$ | 14 | $2.0^{+1.2}_{-0.8}$ | 14 | $3.9^{+3.0}_{-1.7}$ | 14 |
| Ti | $3.1^{+2.0}_{-1.2}$ | 4 | $0.90^{+1.06}_{-0.49}$ | 4 | $1.7^{+1.2}_{-0.7}$ | 4 |
| V  | 3.5±0.5 | 1 | 0.87±0.13 | 1 | 1.9±0.3 | 1 |
| Cr | $2.6^{+1.9}_{-1.1}$ | 2 | $0.62^{+0.26}_{-0.18}$ | 2 | $1.5^{+0.9}_{-0.6}$ | 2 |
| Mn | $2.9^{+0.8}_{-0.6}$ | 3 | $0.99^{+0.72}_{-0.42}$ | 3 | $2.0^{+0.4}_{-0.3}$ | 3 |
| Fe | $1.8^{+1.1}_{-0.7}$ | 15 | $0.47^{+0.30}_{-0.18}$ | 15 | ≡1 | 17 |
| Co | *1.2* | 1 | *0.30* | 1 | *0.64* | 1 |
| Ni | *0.72* | 1 | *0.19* | 1 | *0.40* | 1 |

Sensitivity factors relative to Si were calculated from geometric mean values for data shown in Table 3. For Mg and Fe normalized RSF, individual mineral data were first normalized to these elements, and then geometric mean values were calculated. *n* gives the number of samples measured to calculate the geometric mean values. For Co and Ni, recommended values given in italics are derived from Fe-normalized literature values (Stephan, 2001).

### 3.2 Identification of minerals

One of the main objectives of the Rosetta mission is to characterize the elemental and the mineral compositions of the cometary material. This is also an important objective for COSIMA. As many of the minerals identified in cometary material such as pyroxene and olivine carry the same elemental signal, it will be challenging to differentiate between these classes of minerals. In addition, since the individual mineral grain sizes of cometary material are much smaller than the size of the primary ion beam, mixtures of minerals are measured. Statistical methods like PCA, Corico, KNN, RP, Unscrambler, etc. can be applied to further differentiate between minerals (Engrand et al., 2006; Krüger et al., 2011; Varmuza et al., 2011; Varmuza et al., 2014).

## 5 Summary

COSIMA is a high mass resolution dust analyzer that is able to discriminate the mineral and the organic compounds in the mass spectra of dust particles from comet 67P/Churyumov-Gerasimenko (67P/C-G). To prepare the scientific return of the COSIMA analyses, we have characterized a series of minerals relevant to cometary matter with the reference model of COSIMA on ground. Relative sensitivity factors of elements have been derived from these analyses, expressed as ratios normalized to Si, Mg, and Fe. Using COSIMA, we will thus be able to quantify the major element composition of 67P/C-G cometary grains, normalized to Si, Mg, or Fe.

## Acknowledgements

COSIMA was built by a consortium led by the Max-Planck-Institut für Extraterrestrische Physik, Garching, Germany in collaboration with Laboratoire de Physique et Chimie de l'Environnement et de l'Espace, Orléans, France, Institut d'Astrophysique Spatiale, CNRS/Université Paris Sud, Orsay, France, Finnish Meteorological Institute, Helsinki, Finland, Universität Wuppertal, Wuppertal, Germany, von Hoerner und Sulger GmbH, Schwetzingen, Germany, Universität der Bundeswehr, Neubiberg, Germany, Institut für Physik, Forschungszentrum Seibersdorf, Seibersdorf, Austria, Institut für Weltraumforschung, Österreichische Akademie der Wissenschaften, Graz, Austria and is lead by the Max-Planck-Institut für Sonnensystemforschung, Göttingen, Germany.

The support of the national funding agencies of Germany (DLR, grant 50 QP 1302), France (CNES), Austria, Finland, Sweden (Swedish National Space Board, contract No. 121/11) and the ESA Technical Directorate is gratefully acknowledged. The staff of the national CAMPARIS analytical facility in France (University Pierre et Marie Curie, Paris VI) is acknowledged for helping in the measurement of the mineral compositions by electron microprobe. MT acknowledges support by Klaus Tschira foundation.

We thank the Rosetta Science Ground Segment at ESAC, the Rosetta Mission Operations Centre at ESOC and the Rosetta Project at ESTEC for their outstanding work enabling the science return of the Rosetta Mission.